\documentstyle[12pt]{article}

\catcode`@=12
\begin{document}
\thispagestyle{empty}
\begin{flushright} UCRHEP-T160\\June 1996\
\end{flushright}
\vspace{0.5in}
\begin{center}
{\Large\bf Exotic Decays of Strangelets\footnote{Talk given by E.M. at 
the {\it International Workshop on Strangeness in Hadronic Matter} (Budapest, 
Hungary) May, 1996.} \\}
\vspace{1.8in}
{\bf E. Keith and Ernest Ma\\}
\vspace{0.3in}
{\sl Department of Physics\\}
{\sl University of California\\}
{\sl Riverside, California 92521, USA\\}
\vspace{1.5in}
\end{center}
\begin{abstract}\
The stability of strange matter depends on the implicit 
assumption that baryon number is conserved.  We examine the relevant 
effective operators which allow strangelets to decay by violating the 
conservation of baryon number.  From the experimental lower limit of 
$10^{25}$ years on the stability of nuclei, we find a lower limit of the 
order $10^{6}$ years on the stability of strangelets against such exotic 
decays.
\end{abstract}

\newpage
\baselineskip 24pt
 
\section{Introduction}

The possibility in principle of absolutely stable matter containing strange 
quarks\cite{1,2} has generated a great deal of interest across many 
subfields of physics.  A crucial implicit assumption for the stability of 
strange matter is that baryon number $B$ is exactly conserved.  Of course, 
we know nuclei are stable against $\Delta B \neq 0$ decays.  The best 
mode-independent experimental lower limit to date is $1.6 \times 10^{25}$ 
years.\cite{3}  However, there may be effective $\Delta B \neq 0$ interactions 
which are highly suppressed for nuclei but not for strangelets.  In the 
following we examine such a hypothesis and show that whereas it is possible 
for strangelets to decay by reducing their baryon number, such a lifetime 
should be longer than $10^6$ years.

In this report, we first review briefly the present experimental constraints 
on $\Delta B \neq 0$ interactions.  We then point out the possible 
consequences of an effective $\Delta B = 2$, $\Delta N_s = 4$ operator on 
the stability of strange matter.  We proceed to discuss some possible 
theoretical origins of such an operator, including $R$-parity nonconserving 
terms in supersymmetric extensions of the standard model.  We conclude by 
deriving a phenomenological lower limit of $10^6$ years on the lifetime 
of strangelets against these exotic decays.

\section{Proton Decay}

In order for the proton to decay, there has to be at least one fermion with 
a mass below that of the proton.  Since only leptons have this property, 
the selection rule $\Delta B = 1$, $\Delta L = 1$ is applicable.  The 
most studied decay mode experimentally is 
$p \rightarrow \pi^0 e^+$, 
which requires the effective interaction
\begin{equation}
{\cal H}_{int} \sim {1 \over M^2} (uude).
\end{equation}
Note that because the above effective operator involves 4 fermions, it is of 
dimension 6, hence the interaction must depend on some large mass $M$ 
to the power $-2$.  The fact that\cite{3}
\begin{equation}
\tau_{exp} (p \rightarrow \pi^0 e^+) ~>~ 9 \times 10^{32} {\rm y}
\end{equation}
implies that $M > 10^{16}$ GeV.  This means that proton decay probes 
physics at the grand-unification energy scale.

\section{Neutron-Antineutron Oscillation and Related Processes}

The next simplest class of effective interactions has the selection rule 
$\Delta B = 2$, $\Delta L = 0$.  The most well-known example is of course
\begin{equation}
{\cal H}_{int} \sim {1 \over M^5} (udd)^2,
\end{equation}
which induces neutron-antineutron oscillation and allows a nucleus to decay 
by the annihilation of two of its nucleons.  Note that since 6 fermions are 
now involved, the effective operator is of dimension 9, hence $M$ appears 
to the power $-5$.  This means that for the same level of nuclear stability, 
the lower bound on $M$ will be very much less than $10^{16}$ GeV.

The present best experimental lower limit on the $n - \bar n$ oscillation 
lifetime is\cite{4} $8.6 \times 10^7$s.  Direct search for $NN \rightarrow 
\pi \pi$ decay in iron yields\cite{5} a limit of $6.8 \times 10^{30}$y.  
Although the above two numbers differ by 30 orders of magnitude, it is 
well established by general arguments as well as detailed nuclear model 
calculations that\cite{6} 
\begin{equation}
\tau (n \bar n) = 8.6 \times 10^7 {\rm s} ~\Rightarrow~ \tau (N N \rightarrow 
\pi \pi) \sim 2 \times 10^{31} {\rm y}.
\end{equation}
Hence the two limits are comparable.  To estimate the magnitude of $M$, 
we use
\begin{equation}
\tau (n \bar n) = M^5 |\psi (0)|^{-4},
\end{equation}
where the effective wavefunction at the origin is roughly given by\cite{7}
\begin{equation}
|\psi (0)|^{-2} \sim \pi R^3, ~~~ R \sim 1~ {\rm fm}.
\end{equation}
Hence we obtain $M > 2.4 \times 10^5$ GeV.  Bearing in mind that 
${\cal H}_{int}$ is likely to be suppressed also by products of couplings 
less than unity, this means that the stability 
of nuclei is sensitive to new physics at an energy scale not 
too far above the electroweak scale of $10^2$ GeV. 
It may thus have the hope of future direct experimental exploration.

\section{Effective Interactions Involving Strangeness}

Consider next the effective interaction
\begin{equation}
{\cal H}_{int} \sim {1 \over M^5} (uds)^2.
\end{equation}
This has the selection rule $\Delta B = 2$, $\Delta N_s = 2$ ($\Delta S = 
-2$).  In this case, a nucleus may decay by the process $N N \rightarrow 
K K$\cite{8}.  Although such decay modes have never been observed, the 
mode-independent stability lifetime of $1.6 \times 10^{25}$y mentioned 
already is enough to guarantee that it is negligible.  However, consider 
now
\begin{equation}
{\cal H}_{int} = {1 \over M^5} (uss)^2,
\end{equation}
which has the selection rule
\begin{equation}
\Delta B = 2, ~~~ \Delta N_s = 4.
\end{equation}
To get rid of four units of strangeness, the nucleus must now convert 
two nucleons into four kaons, but that is kinematically impossible. 
Hence the severe constraint from the stability of nuclei does not seem to 
apply here, and the following interesting possibility may occur.

Strangelets with atomic number $A$ (which is of course the same as baryon 
number $B$) and number of strange quarks $N_s$ may now decay into other 
strangelets with two less units of $A$ and one to three less $s$ quarks. 
For example,
\begin{eqnarray}
(A, N_s) &\rightarrow& (A - 2, N_s - 2) + KK, \\ 
(A, N_s) &\rightarrow& (A - 2, N_s - 1) + KKK, ~etc.
\end{eqnarray}
For stable strangelets, model calculations show\cite{9} that $N_s \sim 0.8 A$, 
hence the above decay modes are efficient ways of reducing all would-be 
stable strangelets to those of the smallest $A$.

Unlike nuclei which are most stable for $A$ near that of iron, the 
energy per baryon number of strangelets decreases with increasing $A$. 
This has led to the intriguing speculation that there are stable 
macroscopic lumps of strange matter in the Universe.  On the 
other hand, it is not clear how such matter would form, because there 
are no stable building blocks such as hydrogen and helium which are 
essential for the formation of heavy nuclei.  In any case, the above exotic 
interaction would allow strange matter to dissipate into smaller and 
smaller units, until $A$ becomes too small for the strangelet itself to be 
stable.  A sample calculation by Madsen\cite{9} shows that
\begin{equation}
m_0 (A) - m_0 (A-2) \simeq (1704 + 111 A^{-1/3} + 161 A^{-2/3})~{\rm MeV},
\end{equation}
assuming $m_s = 100$ MeV, and $B^{1/4} = 145$ MeV.  Hence the $KK$ and 
$KKK$ decays would continue until the ground-state mass $m_0$ exceeds the 
condition that the energy per baryon number is less than 930 MeV, which 
happens at around $A = 13$.

\section{Theoretical Prognosis}

If the standard model of quarks and leptons is extended to include 
supersymmetry, the $\Delta B \neq 0$ terms $\lambda_{ijk} u_i^c d_j^c d_k^c$ 
are allowed in the superpotential.  In the above notation, all chiral 
superfields are assumed to be left-handed, hence $q^c$ denotes the left-handed 
charge-conjugated quark, or equivalently the right-handed quark, and the 
subscripts refer to families, {\it i.e.} $u_i$ for $(u,c,t)$ and $d_j$ for 
$(d,s,b)$.  Since all quarks are color triplets and the interaction must 
be a singlet which is antisymmetric in color, the two $d$ quark 
superfields must belong to different families.  (In the minimal supersymmetric 
standard model, these terms are forbidden by the imposition of $R$-parity.  
However, this assumption is not mandatory and there is a vast body of 
recent literature exploring the consequences of $R$-parity nonconservation.)

To generate an effective $(uss)^2$ operator, we need to evaluate the 
one-loop diagram for
\begin{equation}
s s \rightarrow \tilde b \tilde b,
\end{equation}
where $\tilde q$ 
denotes the supersymmetric scalar partner of $q$, and attach the Yukawa 
interactions
\begin{equation}
\lambda_{usb} u^c s^c \tilde b^c
\end{equation}
to the two $\tilde b$'s.  
This is analogous to a recent calculation\cite{10} of the effective 
$(udd)^2$ operator and involves the exchange of the $t$ quark, the $W$ boson, 
and their supersymmetric partners $\tilde t$, and $\tilde w$.  The end result 
is
\begin{equation}
{\cal H}_{int} = {{3 g^4 \lambda^2_{usb} (\tilde A m_b)^2 m_{\tilde w}} \over 
{8 \pi^2 m^4_{\tilde b} m^4_{\tilde b^c}}} |V_{ts}|^2 F(m_t^2, m^2_{\tilde t}, 
m_W^2, m^2_{\tilde w}),
\end{equation}
where the $\tilde b \tilde b^c$ mass term is assumed to be $\tilde A m_b$, 
$V_{ts}$ is the quark-mixing matrix entry for $t$ to $s$ through the $W$ 
boson, and $F$ is a known function of the four internal masses in the loop.
Using the correspondence of $n - \bar n$ oscillation to $NN$ annihilation 
inside a nucleus, we estimate the lifetime of strangelets from the above 
effective interaction assuming $\tilde A = 200$ GeV, $\lambda_{usb} 
< 1$, and both scalar quark masses greater than 200 GeV, to be
\begin{equation}
\tau > 10^{20} {\rm y}.
\end{equation}
This tells us that such contributions are negligible from the supersymmetric 
standard model.

If we extend the supersymmetric standard model to include additional 
particles belonging to the fundamental {\bf 27} representation of $E_6$, 
inspired by superstring theory, then it is possible\cite{11} to obtain 
an effective $(uss)^2$ operator without going through a loop.  Two new 
interactions $u s \tilde h$ and $\tilde h s^c N$ are now possible, 
where $\tilde h$ is a new color-triplet scalar of charge $-1/3$ and 
$N$ is a new neutral color-singlet fermion.  They combine to form an 
effective $M^{-5} (uss)^2$ operator if $N$ is allowed a Majorana mass, 
thereby breaking baryon-number conservation. For the process
\begin{equation}
(A, N_s) \rightarrow (A - 2, N_s - 2) + KK,
\end{equation}
we estimate the lifetime to be\cite{10}
\begin{equation}
\tau \sim {{32 \pi m_N^2} \over {9 \rho_N}} \left[ {M \over {\tilde \Lambda}} 
\right]^{10} \sim 1.2 \times 10^{-28} \left[ {M \over {\tilde \Lambda}} 
\right]^{10} {\rm y},
\end{equation}
where $m_N \sim 1$ GeV, $\rho_N \sim 0.25$ fm$^{-3}$ is the nuclear density, 
and $\tilde \Lambda \sim 0.3$ GeV is the effective interaction energy scale 
corresponding to using Eq.~(6).  If we assume $M = 1$ TeV, then
\begin{equation}
\tau \sim 2 \times 10^7 {\rm y}.
\end{equation}

\section{Lower Limit on the Exotic Decay Lifetime of Strangelets}

Since $\tau$ depends on $M/\tilde \Lambda$ to the power 10 in Eq.~(18), 
it appears that a much shorter lifetime than that of Eq.~(19) is 
theoretically possible.  However, there is a crucial phenomenological 
constraint which we have yet to consider.  Although two nucleons cannot 
annihilate inside a nucleus to produce four kaons, they can make three 
kaons plus a pion.  The effective $(uss)^2$ operator must now be supplemented 
by a weak transition
\begin{equation}
s \rightarrow u + d + \bar u.
\end{equation}
We can compare the effect of this on $N N \rightarrow K K K \pi$ versus 
that of the $(uds)^2$ operator on $N N \rightarrow K K$ discussed in Ref.[10]. 
We estimate the suppression factor to be
\begin{equation}
{{3 \times 10^{-3}} \over {(4 \pi^2)^2}} |V_{us}|^2 {\tilde \Lambda}^4 
G_F^2 \sim 10^{-19},
\end{equation}
where $G_F$ is the Fermi weak coupling constant. 
Since the stability of nuclei is at least $1.6 \times 10^{25}$ years, this 
gives a firm lower limit
\begin{equation}
\tau > 10^6 {\rm y}
\end{equation}
on the lifetime of strangelets against $\Delta B = 2$, $\Delta N_s = 4$ 
decays.

\section{Conclusion and Outlook}

We have pointed out in this report that an effective $(uss)^2$ interaction 
may cause stable strange matter to decay, but the lifetime has a lower 
limit of $10^6$ years.  This result has no bearing on whether stable 
or metastable strangelets can be created and observed in the laboratory, 
but may be important for understanding whether there is stable strange bulk 
matter left in the Universe after the Big Bang and how it should be searched 
for.  For example, instead of the usual radioactivity of unstable nuclei, 
strange matter may be long-lived kaon and pion emitters.  Furthermore, 
if the particles mediating this effective interaction have masses of order 
1 TeV as discussed, then forthcoming future high-energy accelerators such 
as the LHC at CERN will have a chance of confirming or refuting their 
existence.
\vspace{0.3in}
\begin{center} {ACKNOWLEDGEMENT}
\end{center}

The presenter of this report (E.M.) thanks the organizers of 
{\it Strangeness '96} for their great hospitality.  This work was supported 
in part by the U. S. Department of Energy under Grant No. DE-FG03-94ER40837.

\newpage
\bibliographystyle{unsrt} 

\end{document}